\documentstyle[12pt,epsf,epsfig]{article}
\def\qq{\hbox{~{\lower -3pt\hbox{$<$}}\hskip -6pt\raise
-3pt\hbox{${}_\sim$}}}
\newcommand{\br}{{\bf r}}
\newcommand{\beqn}{\begin{equation}}
\newcommand{\eeqn}{\end{equation}}
\newcommand{\barr}[1]{\begin{array}{#1}}
\newcommand{\earr}{\end{array}}
\newcommand{\beqna}{\begin{eqnarray}}
\newcommand{\eeqna}{\end{eqnarray}}

\newcommand{\gapproxeq}{\lower.7ex\hbox{$\;\stackrel{\textstyle>}
{\sim}\;$}}
\newcommand{\plabel}[1]{\label{#1}}
\newcommand{\pbibitem}[1]{\bibitem{#1}}
\begin {document}

\title{
\begin{flushright} 
\small{hep-ph/0009135} \\ 
\small{LA-UR-00-4039} 
\end{flushright} 
\vspace{0.6cm}  
\Large\bf Hadron Structure and Modern Spectroscopy}
\vskip 0.2 in
\author{Philip R. Page\thanks{\small \em E-mail:
prp@lanl.gov}
\\\small \em  Theoretical Division, Los Alamos
National Laboratory, Los Alamos, NM 87545, USA\\
\\ Lectures at the Advanced School 
on Quantum Chromodynamics 2000\\Benasque, Huesca, Spain, 3 - 6 July 2000
}
\date{}
\maketitle
\begin{abstract}{The colour, flavour, spin and $J^{PC}$ of glueballs
and hybrid mesons and baryons are constructed in an intuitive manner in both
the gluon counting and adiabatic definitions. Glueball decay, production
 and mixing and
hybrid meson decay selection rules and production are clarified.
}
\end{abstract}
\bigskip

In the arena of strong nuclear interactions, there are three distinct levels of
understanding.  First there is {\it Quantum Chromodynamics} (QCD), the
Lagrangian of which is relativistic and non-linear (containing three- and four-particle
interactions).  Because the lowest energy state (vacuum) is
non-empty, it can be thought of as a many-body system of particles. 
In addition,
QCD is a quantum field theory.  All these features conspire to make 
the physical
predictions in the regime of strong interactions largely intractable.  There
hence exists a second level of understanding, called {\it phenomenology},
which attempts to capture strong interaction phenomena by use of simplified
pictures.  Phenomenology receives ``data'' from the first level by 
virtue of QCD and its 
computational expression, called lattice QCD.  Phenomenology 
also receives
data from the third level of understanding: {\it experiment} on 
strongly interacting particles (hadrons), sometimes
called ``empirical hadron spectroscopy''.  These lectures concern the 
data stream
between levels two and three.  Phenomenology serves to guide and interpret
calculations on level one, and observations on level three.
It does not make a claim to precision, and that proviso should be kept 
in mind throughout our discussion.

However, these noble features are not enough to protect level two from
extinction, and replacement by a sole data stream between levels one and three.
So why do we study phenomenology?  Ultimately it is because it provides a
language in which to express strong interactions, so that the phenomenon can be
comprehended by the human mind.

These lectures will present highlights on explicit excitations of the force
carriers, i.e., the gluons. A bibliography of recent books and reviews is provided for further reference.

Towards this purpose, we briefly review the non-relativistic quark model of
quark-antiquark pairs (mesons) and three-quark composites (baryons).

\section{Quark Model}

A meson at rest can be represented by
\beqn\plabel{mes}
\psi({\bf r}) \delta_{c\bar{c}} {\cal F}_{f\overline{f}} {\cal S}_{s\bar{s}} q^+_{cfs}
(\frac{\bar{m}\br}{m + \bar{m}})\bar{q}^{\; +}_{\bar{c}\bar{f}\bar{s}}
(\frac{-m \bar{{\bf r}} }{m + \bar{m}}) | 0 \rangle
\eeqn
Here implicit summation or integration over respectively discrete 
(subscripted) and continuous
variables is implied.

The labels ${\bf r}$, $c$, $f$, $s$ and $m$ denote the position, 
colour, flavour,
(non-relativistic) spin and mass of the quark, which is created by 
the operator $q^+$ from the
vacuum $|0\rangle$. Accordingly for the antiquark.
The spacial wave function is $\psi$, and the flavour and
spin structures $\cal F$ and $\cal S$ respectively.  The 
orbital angular momentum
$\bf L$ is conserved with $\psi$ carrying the quantum 
numbers $L$,
$L_z$.  The spin $\bf S$ is conserved with $\cal S$ carrying 
the quantum numbers
$S$, $S_z$.  The total angular momentum ${\bf J} = {\bf L} + {\bf S}$ is
also conserved, with the meson carrying quantum numbers $J$, $J_z$.
The coefficient that expresses this is suppressed in Eq. \ref{mes}.

Under reflection through the origin (parity), ${\bf r} \rightarrow - {\bf r}$
in $q^+
\bar{q}^{\,+}$, and an additional sign appears because the intrinsic 
parity of an antiquark
is opposite to that of a quark.  The latter property holds for 
fermions in field theory,
{\em i.e.}, comes from the first level of understanding.  Noting that
$\psi(-{\bf r}) = (-1)^L
\psi({\bf r})$, the parity $P = (-1)^{L+1}$.  

Particle-antiparticle exchange (charge
conjugation) interchanges $q^+$ and $\bar{q}^{\,+}$.  Assume for the 
purpose of this paragraph that the quark and the antiquark 
have the same flavour.
Noting that fermionic creation operators anticommute; that if the 
quark and antiquark have
the same flavour, $F_{\bar{f}f} = F_{f\bar{f}}$; and that 
$S_{\bar{s}s} = (-1)^{S+1}
S_{s\bar{s}}$, one can conclude that the charge conjugation $C = 
(-1)^{L + 1 + S + 1} = (-1)^{L + S}$.  It follows that $CP = (-1)^{S+1}$.

Given the equations ${\bf J} = {\bf L} + {\bf S}$, $P = (-1)^{L + 1}$ and 
$C = (-1)^{L + S}$ one 
can construct the
$J^{PC}$ of all mesons.  It can then be checked that the combinations 
$J^{PC} = 0^{--}$,
$0^{+ -}$, $1^{- +}$, $2^{+ -}$, $3^{-+} \ldots$ are not allowed. 
These will be referred to
as ``$J^{PC}$ exotic''.

When the up $u$ and down $d$ quarks
 can be treated the same, {\em i.e.,} when their
different electric charges and masses are neglected, the isospin $\bf I$ is
conserved with ${\cal F}$ carrying the quantum numbers $I, I_z$.  $G$-parity is
defined as two operations conjoined:  charge conjugation and a 180$^o$ degree
rotation is isospin space.  The latter is equivalent to the transformation
$u \rightarrow d$, $d \rightarrow - u$.  
Consider an isospin multiplet built only from $u$ and $d$ quarks.
It is possible to show that all
states in the multiplet carry the same quantum number $G$
\cite{close79}.  Consider
the $I_z = 0$ member of the multiplet, which also carries the quantum
number $C$.  By the definition of $G$-parity, $G = (-1)^IC$, where under $u
\rightarrow d$, $d \rightarrow -u$, ${\cal F}_{f\bar{f}} \rightarrow (-1)^I
{\cal F}_{f\bar{f}}$.

Baryons can be constructed via three quark creation operators, as a
straight-forward variation of the meson case, except that the colour changes
from the Kronecker delta $\delta_{c\bar{c}}$ to the 
(totally antisymmetric) $\epsilon$-tensor $\epsilon_{c_1c_2c_3}$.
All the quantum numbers remain conserved, except for charge conjugation, which
changes a baryon into an antibaryon and hence does not correspond to a
quantum number.  
Because charge conjugation does not correspond to a quantum number, the same
is true for the derivative operation $G$-parity.
One can show by enumerating all possibilities that all $J^P$
are possible for baryons, so that there are no exotic $J^P$.
  The Kronecker delta and
$\epsilon$-tensor are the only tensors available in the fundamental
representation of colour $SU(3)$ in which quarks live in QCD
\cite{close79}.  They are both
employed in such a way as to force the meson or baryon to carry no colour
labels, {\em i.e.,} to be white.  This requirement arises from the third level
of understanding, called {\it confinement} \cite{simonov99}:
  Since no free (colour carrying) quarks
or gluons have ever been observed, all free particles are taken to be white.

\section{New White Particles}

In 1972 Murray Gell-Mann and Harald Fritzsch 
realized that there is a zoo of new white particles, among them:

{\bf Glueballs}:  The colour structure $\delta _{\gamma_1 \gamma_2}$, in 
the adjoint
representation of colour $SU(3)$ in which gluons live in QCD, is 
overall white for {\em two
gluons}. The colour structures given by the invariant $SU(3)$ 
tensors
$f_{\gamma_1\gamma_2\gamma_3}$ and $d_{\gamma_1\gamma_2\gamma_3}$
in the adjoint representation are overall white for {\it three gluons}
\cite{close79}.

{\bf Hybrid mesons}:  The colour structure 
$\lambda^\gamma_{c\bar{c}}$ is overall
white for a {\em quark, antiquark and a gluon}, where $\lambda$ is a 
Gell-Mann matrix \cite{close79}.

{\bf Hybrid baryons}:  The colour structure $\lambda^\gamma_{c_1 c^\prime}$
$\epsilon_{c^\prime c_2 c_3}$ is overall white for {\em three quarks 
and a gluon}.

{\bf Four-quark states or ``Meson molecules''}:  The colour structures
 $\lambda^\gamma_{c_1 \bar{c}_1} \lambda^\gamma_{c_2
\bar{c}_2}$ and  $\delta_{c_1\bar{c}_1}
\delta_{c_2 \bar{c}_2}$ are overall white for {\em two quarks and two antiquarks}.

These definitions of a glueball, hybrid meson and baryon, where we
have a specific number of gluons, will be referred to as {\it gluon counting}.

Glueballs, being composed only out of gluons, cannot carry any 
flavour, as this is a
property of quarks.  Particularly, this implies that they have $I = 
I_z = 0$.  Hybrid
mesons and baryons are respectively mesons and baryons with an 
additional gluon, so they
have the same flavour structure.  Four-quark states have a more 
complicated flavour
structure.

Now we summarize some properties that follow from the 
first level of understanding. 
A gluon field has $J = 1$, which means that it is a four-vector with
both a time-like and three space-like components. The time-like component has
$P=1$ and the space-like components $P=-1$. However, not all these
components are dynamical. This is because QCD is invariant under
local $SU(3)$ colour transformations, which transform the gluon field.
Because all these transforms of the gluon field are equivalent, one uses
only one version, called {\it gauge fixing}. This restricts the
gluon field to have only three dynamical components. These can be thought of
as the space-like components with $P=-1$. The photon field, which mediates the
electromagnetic interaction, has identical properties. In addition it
also has $C=-1$. Accordingly, 
it has been verified experimentally in electron-positron ($e^-e^+$) annihilation into
a photon that the photon field has 
$J^{PC} = 1^{--}$. The charge conjugation for the gluon field 
is not so simple.
This is because a blue-antired gluon would for example transform to a
red-antiblue gluon. We shall loosely say that the gluon has $C=-1$,
although there will be exceptions.

In free space a gluon can have a continuous range of momenta.
When one puts the gluon inside an enclosure its momenta become discrete.
The lowest two momenta are called ``magnetic'' (also called 
``transverse electric'',
TE), and ``electric'' (also 
called ``transverse
magnetic'', TM).  TE 
gluons have $J^{PC} =
1^{+-}$ and TM gluons $1^{--}$.

Let's build the $J^{PC}$ of our new white particles.

{\bf Glueballs}:  Two gluons together will 
hence 
have $J^{PC} =
(0, 1, 2)^{++}$ when they have no orbital angular momentum relative 
to each other, called
{\it $S$-wave}.
With one unit of angular momentum relative to each other, called 
{\it $P$-wave}, corresponding to
higher mass particles, the glueballs will have $J^{PC} = (0, 1, 
2)^{++} \otimes 1^- = (0,
1, 2, 3)^{-+}$.  

Since the first level of understanding states that 
gluons are massless
before any interactions, and using the {\it Yang-Landau theorem} that massless $J = 
1$ particles do not
couple to two identical massless $J = 1$ particles \cite{close79}, 
we deduce that $J = 1$ 
glueballs are not
allowed. Because the gluons are not massless after interactions one 
expects that the $J = 1$ 
glueballs would have a substantial mass. This is confirmed by lattice QCD
\cite{sharpe99,richards00}.
Hence the lightest glueballs are expected to be $0^{++}$ 
and $2^{++}$, with the
next lightest $0^{-+}$, $2^{-+}$ and $3^{-+}$.  This mass ordering is 
confirmed by lattice
QCD \cite{sharpe99,richards00}.
Some three-gluon composites have $C = -$ since there are an odd 
number of gluons.
Because gluons do have some mass due to self-energy, these are 
expected to be heavier than
the lowest two-gluon glueballs.  This is indeed found in lattice QCD
 \cite{sharpe99,richards00}.

{\bf Hybrid Mesons}:  The $J^{PC}$ can be obtained by adding the 
$J^{PC}$ of the
lowest lying  quark-antiquark composites in the quark model, $0^{-+}$ 
and $1^{--}$,
corresponding to $S = 0$ and $1$ respectively, to the $J^{PC}$ of the 
gluon.  For TE
gluons, this gives $(0^{-+}, 1^{--}) \otimes 1^{+-} = 1^{--}, (0, 1, 2)^{-+}$.
One immediately notes that $1^{--}$, $0^{-+}$ and $2^{-+}$ have the 
opposite spin
assignment $S$ to what they would have if they were mesons.  The 
remaining $S=1$ state $1^{-+}$
is $J^{PC}$ exotic.  

For TM gluons, which are heavier than TE gluons 
in bag models \cite{close79},
the low-lying hybrids have $J^{PC} = (0^{-+}, 1^{--}) \otimes 1^{--} 
= 1^{+-}, (0, 1,
2)^{++}$.  These are identical to the $L = 1$ mesons, with the same 
spin assignments.  

We hence expect the lightest $J^{PC}$ exotic hybrid to be $1^{-+}$, 
which is confirmed by
lattice QCD \cite{barnes97}.

{\bf Hybrid Baryons}:  One may think that 
the $J^P$ is found by adding the $J^P$ of the 
low-lying
three-quark composites, $\frac{1}{2}^+$ and $\frac{3}{2}^+$, 
corresponding to $S =
\frac{1}{2}$ and $\frac{3}{2}$ respectively, to the $J^P$ of the 
gluon.  For TE gluons,
this gives $\left(N\frac{1}{2}^+, \Delta \frac{3}{2}^+ \right) \otimes 1^+ = N
\left(\frac{1}{2}, \frac{3}{2}\right)^+$, $\Delta \left(\frac{1}{2}, 
\frac{3}{2},
\frac{5}{2} \right)^+$.

More careful study, including constraints from the Pauli Principle 
that two fermions
(quarks) cannot occupy the same state, implies that the $S=\frac{1}{2}$ 
hybrid baryons are $N \left(\frac{1}{2}, \frac{3}{2}\right)^+$ and
$\Delta \left(\frac{1}{2}, \frac{3}{2}\right)^+$, and the $S=\frac{3}{2}$ 
hybrid baryons are $N \left(\frac{1}{2}, \frac{3}{2},
\frac{5}{2} \right)^+$, so that there are seven 
low-lying TE hybrid baryons \cite{barnes93}.

A TM gluon has the same quantum numbers as a TE one, except
for parity. The quantum numbers of the TM hybrid baryons are 
accordingly identical to
the previous paragraph, except that all states have $P=-$. 

{\bf Four-quark states or ``Meson Molecules''}:  By looking at 
composites of two
mesons, with some orbital angular momentum between them, it is easily 
shown that all
$J^{PC}$ are principle allowed.

The main feature of four-quark states is that they can fall apart 
into two mesons without
inhibition, by simply arranging their colour structure to that of two mesons.
One should hence regard them as being too unstable to be observed in 
experiment unless
specific dynamics dictate otherwise.

The above definitions of glueballs, hybrid mesons and baryons relied 
on the notion that the
gluons can be enumerated.  However, this is by no means clear, as 
non-interacting gluons
are massless, which would make stochastic multigluonic configurations just as 
massive as the cases
listed so far.  An alternative approach is suggested by fixing the 
positions of all the
quarks and antiquarks and calculating the energy of the 
system, called the {\em
adiabatic potential}, as a function of quark/antiquark positions. 
Because QCD is a quantum
theory, there will not only be a ground state adiabatic potential but 
also
excited adiabatic potentials.  Allowing the quarks and antiquarks to 
be heavy but not fixed
may conceivably allow the following {\em adiabatic approximation}.

First calculate the adiabatic potentials by fixing the quark and 
antiquark positions.  Then
allow the heavy quarks and antiquarks to move in the adiabatic 
potentials just calculated.
If the masses thus obtained are identical to masses from first 
principles, we say that the
adiabatic approximation is valid.  This is dependent on whether the 
quarks and antiquarks
can be regarded as moving slowly with respect to the gluons.

If the adiabatic approximation is valid, as can be shown for a
quark-antiquark \cite{isgur90} or three
quarks moving on the ground state adiabatic potential, one can {\em 
define} these systems
as mesons or baryons respectively.  The $J^{PC}$ of the ground state
potential is 
$0^{++}$, as verified
by lattice calculations \cite{sharpe99}. 
Such a potential will not change the quantum 
numbers previously
calculated for mesons and baryons.  If the adiabatic approximation is 
valid for the low-lying excited
adiabatic potential, one can {\em define} the low-lying hybrid
mesons or baryons 
as a quark/antiquark
or respectively, three-quarks, moving in this 
potential. This is referred to as the {\it adiabatic}
definition.

{\bf Hybrid Mesons}:  When one fixes the quark and the antiquark it 
is clear that the
system is invariant under rotations around the line between the 
quark and the antiquark (see problem \ref{symm}).
If the orbital angular momentum around this line is $\Lambda$, one can 
form degenerate
states $|\Lambda \rangle$ and $|- \Lambda\rangle$.
These states are degenerate since the energy cannot depend on whether the
system rotates clockwise or anticlockwise. Any linear combination of
$|\Lambda \rangle$ and $|- \Lambda\rangle$ has the same energy.
The action of parity is to 
interchange $|\Lambda
\rangle$ and $|- \Lambda \rangle$, since it interchanges clockwise and 
anticlockwise rotations.  The
same is true for charge conjugation, which interchanges the quark and
antiquark, i.e. changes the direction of
the rotation axis, and hence makes clockwise rotations anticlockwise, and
vice versa. 
One can now construct the
eigenstates of parity and change conjugation $\frac{1}{2} 
(|\Lambda \rangle \pm  | - \Lambda \rangle
)$.

Taking from lattice QCD that the potential has $|\Lambda| = 1$ and $C = 
-P$ \cite{sharpe99}, and using the eigenstates above, it follows that
the $J^{PC}$ of the adiabatic potential is $1^{+-}$ 
or $1^{-+}$.  
Technically $J$ is not a quantum number of the adiabatic potential, but
only $|\Lambda|$ (see problem \ref{symm}). We loosely equate
$J$ and $|\Lambda|$. 
The low-lying hybrid mesons are $(0^{-+}, 1^{--}) \otimes (1^{+-}, 
1^{-+})$ $= 1^{--}, (0,
1, 2)^{-+}, 1^{++}, (0, 1, 2)^{+-}$.
There is the same number of states as in the previous definition of a 
hybrid meson, with
six having the same $J^{PC}$.  Note that all non-exotic $J^{PC}$ 
adiabatic hybrids have the opposite
spin $S$ than what they would have if they were conventional mesons. 
The states $1^{-+},
0^{+-}$ and $2^{+-}$ are $J^{PC}$ exotic.  Lattice QCD confirms that 
these are the three 
lightest $J^{PC}$
exotic hybrids \cite{barnes97}.

Within the adiabatic definition of a hybrid, it is possible to specialize
to the case of gluon counting, so that the two definitions do not have
to be disjoint. An example is the adiabatic bag model where the hybrid is
still defined as a quark-antiquark-gluon composite but studied using the
adiabatic approximation. One finds that the TE hybrids have the same 
quantum numbers as outlined for adiabatic hybrids in the previous paragraph.
There are hence eight of them, in contrast to the four TE hybrids originally
discussed in the gluon counting definition!

{\bf Hybrid Baryons:} The Isgur-Paton flux-tube model 
\cite{paton90,paton98} indications are
that the low-lying
excited adiabatic potential has $J^{PC} = 1^{++}$. This yields five
hybrid baryons with $J^P = (N\frac{1}{2}^+,\Delta\frac{3}{2}^+) \otimes
1^+$ = $N(\frac{1}{2},\frac{3}{2})^+ , \Delta(\frac{1}{2},\frac{3}{2},
\frac{5}{2})^+$, with the former two states having spin $\frac{1}{2}$,
just like the conventional $N$, and the latter three states having spin
$\frac{3}{2}$, just like the conventional $\Delta$ \cite{capstick00}.
The reason why the
Pauli Principle does not change this simple argument is that the
quark label exchange properties of the colour structure remain totally
antisymmetric for at least some
hybrid baryons in the  flux-tube model, as it is for
the $\epsilon$-tensor of conventional baryons. Note that
four of the five low-lying hybrid baryons agree, as far was their
flavour and $J^P$ are concerned, with the seven low-lying TE hybrid baryons
according to the former definition. However, when spin $S$ is
considered in addition, this is only true for two of the five hybrid baryons.

What about an adiabatic definition of glueballs? Conceptually this is
difficult because there are no heavy quarks that can be treated as
moving adiabatically. Hence only hybrid mesons and baryons and 
four-quark states can possibly be
described by the adiabatic definition.

The way glueballs, conventional and
hybrid meson and baryons, and four-quark states
were described sofar did not allow for the possibility of mixing between
different types of states with the same quantum numbers $J^{PC}$ or $J^P$. The
unmixed states are referred to as {\it primitive} (bare), and the mixed states
as {\it physical} (dressed).

\section{Decay \plabel{selsec} }

There is always the possibility that gluons
will allow a quark-antiquark pair to be created, called {\it decay},
coming from the first
level of understanding. 

If initial state $A$ decays to final states $B$ and $C$, several quantum 
numbers are conserved. A straightforward example is the
electric charge. For total angular momenta,
${\bf J}_A ={\bf J}_B + {\bf J}_C
+ \tilde{\bf L}$, where $ \tilde{\bf L}$ is the relative orbital angular
momentum between $B$ and $C$. Also, for parity, 
$P_A = (-1)^{\tilde{L}} P_B P_C$.
When all the states have well-defined $C$,
charge conjugation conservation gives $C_A = C_B C_C$. For isospin
symmetry, ${\bf I}_A = {\bf I}_B + {\bf I}_C$. For all states having
well-defined G-parity $G$, $G_A = G_B G_C$.

I shall now discuss what is qualitatively
known about decays of glueballs and hybrid mesons. Little is known
about the decays of hybrid baryons and four-quark states.

{\bf Glueballs}: Glueballs, in the limit where the $u, d$ and
strange $s$
flavour quark behave the same, called $SU(3)$ {\it flavour} symmetry, are
expected to decay to the $\pi$, $\eta$ and $\eta'$ as follows. We
respectively use the $SU(3)$ flavour structures
$\frac{1}{\sqrt{2}}(u\bar{u} - d\bar{d}),\;
\frac{1}{\sqrt{6}}(u\bar{u} + d\bar{d} - 2 s\bar{s})$ and
$\frac{1}{\sqrt{3}}(u\bar{u} + d\bar{d} +s\bar{s})$. Then

\begin{table}[t]\centering
\begin{center}
\begin{tabular}{|c|c|c|c|} \hline
& Amplitude & Width & Final states  \\
\hline
$G \rightarrow \pi\pi$ & 1 & 3 & $\pi^+ \pi^-, \pi^- \pi^+, \pi^0 \pi^0$  \\
$G \rightarrow K \bar{K}$ & 1 & 4 & $K^+ K^-, K^- K^+, K^0 \bar{K}^0, 
\bar{K}^0 K^0$  \\
$G \rightarrow \eta \eta$ & 1 & 1 & $\eta \eta$  \\
$G \rightarrow \eta^\prime \eta$ & 0 & 0 & $\eta^\prime \eta, \eta 
\eta^\prime $  \\
\hline
\end{tabular}
\end{center}
\caption{\plabel{demo}
Ratios of intrinsic amplitudes to one final state, and
widths to all final states.}
\end{table}

\begin{eqnarray}\plabel{gdec}
\left <G|\pi^+\pi^- \right > &  = & \left < 0|u \bar{d} d \bar{u} 
\right > = 1 = \left <
G|\pi^0 \pi^0 \right > \nonumber\\
\left <G| K^+ K^- \right > & = & \left < 0|u \bar{s} s \bar{u} \right > = 1 =
\left <G| K^0 \bar{K}^0 \right > \nonumber\\
\left <G|\eta \eta \right > & = & \left <  0|\frac{1}{\sqrt{6}}
\left(u \bar{u}  + d \bar{d} - 2 s \bar{s} \right ) \frac{1}{\sqrt{6}}
\left(u \bar{u}  + d \bar{d} - 2 s \bar{s} \right ) \right >
    =  \frac{1}{6} (1 + 1 + 4) = 1 \nonumber\\
\left < G| \eta' \eta \right > & = & 0
\end{eqnarray}
This decay pattern is indicated in Table \ref{demo} and is
called {\it flavour democratic} decay. The decay topology assumed for
glueball decay is topology 4a in Fig. \ref{top}. This is called
an {\it Okubo-Zweig-Iizuka (OZI) forbidden} decay, because the ``half-doughnut''
final state can be ``pulled away'' from the initial glueball, i.e. it 
is possible to cut through the topology without intersecting a quark line. 
Topology
4b is {\it double OZI forbidden}, because both of the final ``raindrops'' can
be pulled away separately from the glueball. The 
(phenomenological) OZI rule states that the size of decay decreases as
the number of components in a topology that can be pulled away from 
each other increases \cite{leyaouanc88}.

Flavour democratic decay was not confirmed in lattice QCD in the
$SU(3)$ limit \cite{richards00}. 
This invalidates the intuitive argument presented above.
 From a heuristic
point of view, glueball decay includes two possibilities:
Firstly, the glueball can decay directly to two mesons, in the sense that
the two quark-antiquark pairs are created at a similar time.
Secondly, the glueball can mix with a meson, and the meson then decays
at a later time to two mesons. Here the idea is that the first
quark-antiquark pair is created long before the second.
The first possibility is called {\it primitive glueball decay}, while the
second is due to {\it glueball-meson mixing}. Although it is not possible to
rigorously separate these two notions, current modelling suggests that
glueball-meson mixing can explain the lattice results without the
need to invoke primitive glueball decay.

\begin{figure}
\begin{center}
\leavevmode
\hspace{-.5cm}\hbox{\epsfxsize=5 in}
\epsfbox{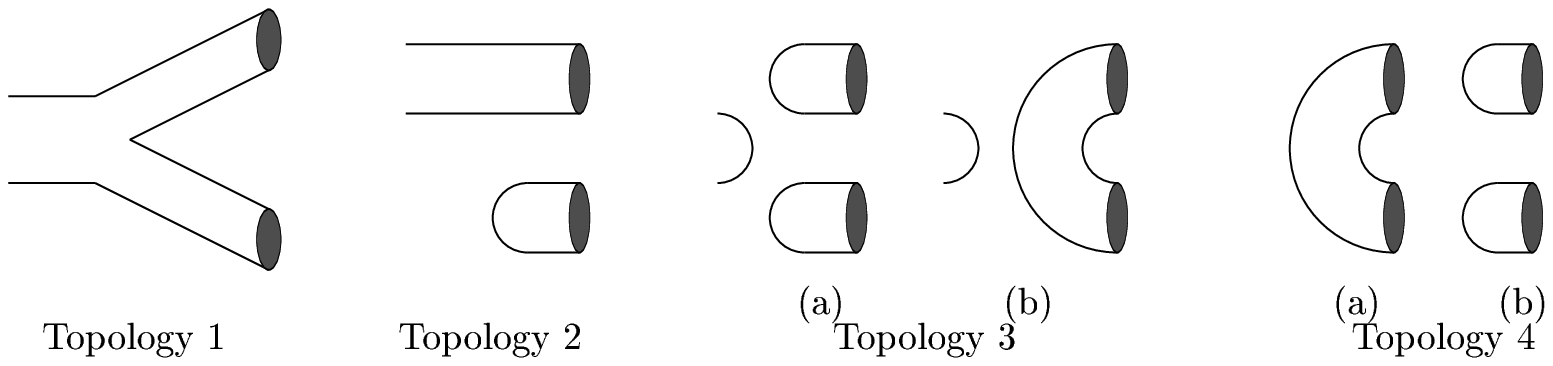}
\vspace{-1.2cm}\caption{\plabel{top}
Decay topologies.}\end{center}
\end{figure}

{\bf Hybrids:}
Consider topology 1 in
Fig. \ref{top}. Each of the three participating
quark-antiquark pairs are connected to each other, called {\it connected} 
decay. None can be pulled away from the other, i.e. the decay is {\it OZI allowed}  
and hence expected to be dominant. Note that
the quark in the initial state ends up in the one final meson,
and the antiquark in the other meson. Topologies 2 and 3b are
single OZI forbidden, and topology 3a double OZI forbidden.

Let us explore connected decay within the adiabatic definition of conventional
and hybrid mesons. Under the adiabatic approximation one can fix
the positions of all quarks and antiquarks participating in the decay.
This means that there must be an amplitude for the gluons of the initial
state to fold into the gluons of the two final states. This
{\it flux-tube overlap} depends on the variables that specify the
configuration: the distance between the initial quark and antiquark,
and the vector from the midpoint of the  initial quark-antiquark 
line to the pair creation position (see problem \ref{flux}). The
spacial orientation of the initial quark-antiquark line is irrelevant by
rotational invariance. A flux-tube overlap will also exist for the
decay of conventional or hybrid baryons.

Consider connected decay. Assume that the quark-antiquark pair
 creation is with
spin $\tilde{S}=1$. Then we deduce the following {\it spin selection rule}:
Spin $S_A=0$ mesons do not decay into two spin $S_B=S_C=0$ mesons.
This follows simply because the total spin in the initial state is
$0$, while the total spin in the final state is $1$, because  $\tilde{S}=1$,
so that spin is not conserved in the decay. This selection rule holds
spectacularly better for conventional meson decay than one may expect. 
As recently measured by VES, the
decay $\pi_2(1670) \rightarrow b_1\pi$, 
where each participating meson
is spin 0, has a minute branching ratio of less than
$0.2\%$ at the $2\sigma$ confidence level. 

Assuming the spin selection rule to also be valid for hybrid meson
decay, one obtains important experimental implications. It has already been
pointed out before that the low-lying non-exotic TE hybrid in the gluon 
counting definition, and all the low-lying non-exotic
hybrids in the adiabatic definition, have the opposite spin assignment 
than their conventional meson partners with the same $J^{PC}$.
Restrict the discussion of hybrid mesons in this paragraph to these
hybrids. 
Consider a decay of an initial state to two final states
where the spin selection rule is operative. Then it follows that if the
nature of the initial state is interchanged between a conventional and
a hybrid meson, the spin selection rule will no longer be valid.
For example, if $\pi_2(1670)$ was a hybrid meson, its decay to
$b_1\pi$ would be uninhibited.
This means that the conventional or hybrid meson nature of the
initial state can be distinguished
based on whether the width is suppressed or not.

There are two further selection rules which are more general than 
specific models:

\begin{figure}[t]
\begin{center}
\vspace{-1.3cm}
\hspace{-0.5cm}\epsfig{file=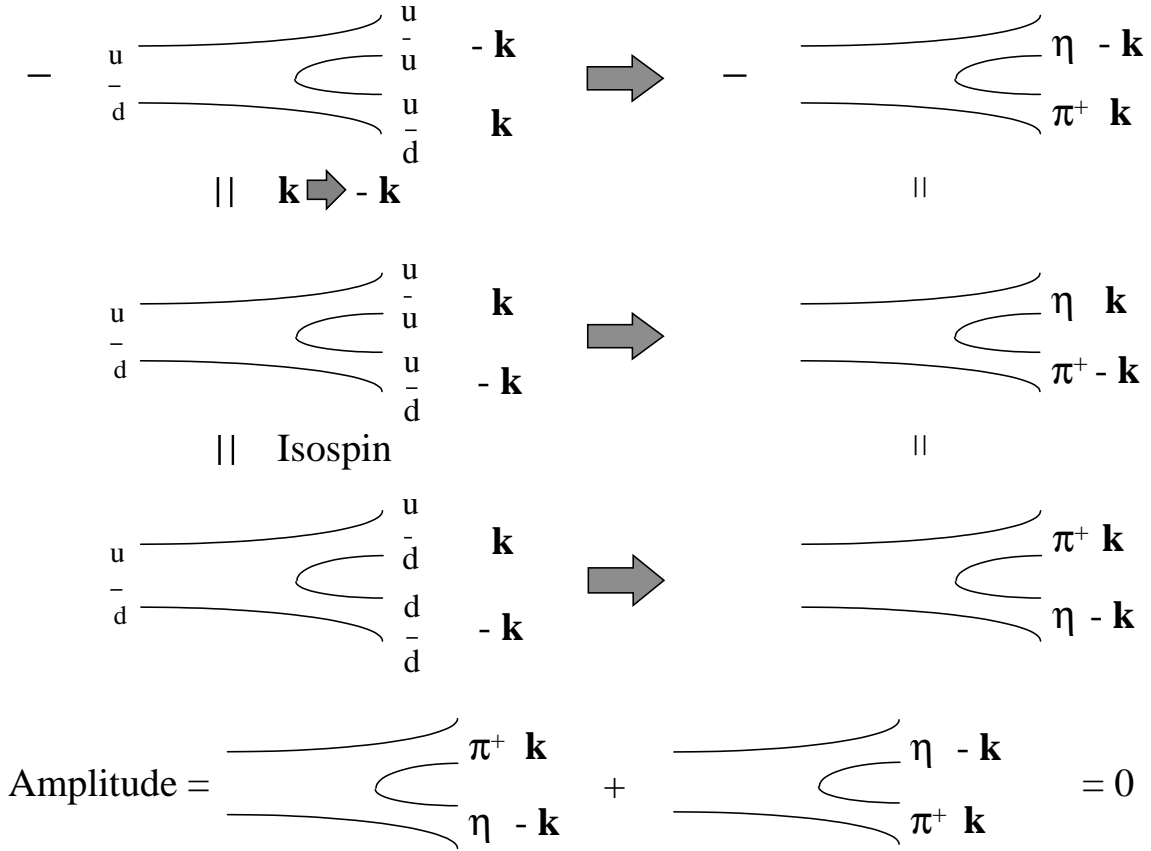,width=17cm,angle=0}
\vspace{-1cm}
\caption{\plabel{lip} Selection Rule I.}
\end{center}
\end{figure}

{\bf I}.
{\it $J^{PC} = 1^{-+},\; 3^{-+},\; \ldots$ flavour structure $q\bar{q}$
hybrid mesons do not proceed via connected 
decay to $\eta\pi$.} Here $q\bar{q}$ refers to the initial state
having the same flavour quarks and antiquarks. If isospin symmetry
is assumed for a decay involving only $u,d$ quarks, the result
can be extended to all members of the isospin multiplet.
This rule, originally noticed by Lipkin, LeYaouanc, Oliver,
P\`{e}ne and Raynal in 1988-89, does not follow from any standard 
conservation 
principle, and is specific to the connected topology, in the sense that
it is known not to be valid for topology 2. The derivation does not
depend on assuming non-relativistic behaviour, and can in fact be
derived from the first level of understanding. 

We outline an intuitive derivation for the decay of a positively charged
$J^{P} = 1^{-},\; 3^{-},\; \ldots$ meson to $\eta\pi^+$ when
isospin symmetry is assumed. This decay is allowed by all the 
conservation principles listed in the beginning of this section.
Because G-parity conservation implies that
the neutral isospin partner of the initial state is $J^{PC}$ exotic, the
initial state must be a hybrid meson. The gluons in the
connected decay (topology 1 in Fig. \ref{top}) are not indicated. 
The argument is depicted in Fig. \ref{lip}. Taking the initial hybrid
at rest, the $\eta$ and $\pi^+$ emerge with momenta $-\bf k$ and 
$\bf k$ respectively. First consider the three top left diagrams. The top
diagram has a negative sign in front by convention.
When the transformation ${\bf k} \leftrightarrow -{\bf k}$ is applied,
the middle diagram is obtained, noting that an odd $\tilde{L}$ decay 
acquires an extra minus sign. This is a general property of
odd $\tilde{L}$ decays. The bottom diagram is obtained by noting that
the amplitude to create a $u\bar{u}$ pair is the same as for a
$d\bar{d}$ pair within isospin symmetry. 
The three top right diagrams are now obtained from the
three top left (quark) diagrams by attaching the initial hybrid to the
initial $u\bar{d}$ quarks, and the final $\pi^+$ to the final 
$u\bar{d}$ quarks. Since the flavour wave function of the $\eta$ is
proportional to $u\bar{u}+d\bar{d}$, it is attached to either $u\bar{u}$
or $d\bar{d}$, with a positive relative sign. Because each of the
three top left diagrams are equal, it follows that each of the three
top right diagrams are equal. The bottom diagrams depict the decay
amplitude, taking into account that there are two possible ways for the
final $\eta$ and $\pi^+$ to couple. Looking back at the
top right diagrams one immediately notices that the decay amplitude 
vanishes. This is the selection rule.

{\bf II}.
{\it  Flavour structure $q\bar{q}$ hybrid mesons
do not proceed 
via connected decay to 
two $L_B=L_C=0$ conventional 
mesons which are identical, except possibly for their flavour
and spin, under $\tilde{S}=1$ quark-antiquark pair creation}
\cite{barnes97,godfrey99,page95}. 
Here restrict the hybrid mesons to the four low-lying TE hybrids in the gluon 
counting definition, and all the eight low-lying 
hybrids in the adiabatic definition. Evidently, non-relativistic
behaviour is assumed. 
The same comments about isospin symmetry made for the first rule apply
here.

The general derivation of this rule is somewhat complicated, but a simple
derivation obtains for hybrids in the adiabatic definition {\it if}
the following ansatz is made: the $CP$ of the participating 
adiabatic potentials and the $CP$ of the created pair are conserved\footnote{The general derivation is in P.R. Page, {\it Phys. Lett.} {\bf B402} (1997) 183, and the ansatz in C. Michael, {\it 8$^{th}$ Int. Symp. on Heavy Flavor Physics (Heavy Flavors 8)}, Southampton, UK, 25-29 July 1999; hep-ph/9911219.}.  
The ansatz means that $-1 = 1 \times 1 \times (-1)^{\tilde{S}+1}$:
We used that the hybrid and conventional meson potentials have respectively
negative and positive $CP$, and the created pair has $CP=(-1)^{\tilde{S}+1}$.
When $\tilde{S}=1$, the ansatz is not satisfied and the decay vanishes:
thus the selection rule.

\section{Production} 

In Fig. \ref{prod} we indicate the main production mechanisms
relevant to spectroscopy. These are $\psi$ (charm-anticharm, $c\bar{c}$)
radiative decay,
proton-antiproton ($p\bar{p}$)
 annihilation, central and diffractive production,
pion ($\pi$) and photon ($\gamma$)
beams, two-photon production and $e^-e^+$ annihilation.  
Processes not listed that have yielded spectroscopical information
include Primakoff production (an incoming particle
in an electromagnetic field), jets, $\tau,\; D,\; D_s, \; B$ and 
$\psi$ hadronic decay, and $K$ beams.
Current experiments are also listed: BES  at the Beijing 
Electron-Positron Collider, CBAR (Crystal Barrel) and  Obelix at
the Low Energy Antiproton Ring at CERN, WA102 and LEP2
at CERN, VES at
Serpukhov, E852 at the Alternating Gradient Synchrotron at 
Brookhaven, Hall B at Jefferson Lab, CLEO
at the Cornell Electron Storage Ring and ARGUS at DORIS II at the
Deutsche Electronen Synchrotron. Only the production of glueballs and 
hybrids are indicated in Fig. \ref{prod}.
The $f_0(1500),\; f_J(1710)$ and $f_J(2220)$ are glueball candidates
and the remainder of the states listed are hybrid meson candidates.

For the cross-sections of various production mechanisms, 
we perform a na\"{\i}ve counting in relative powers of the strong coupling
constant $\alpha_S$ for light quarks.
The first three processes in Fig. \ref{prod} are glue-rich: they
prefer to produce glueballs, with hybrids suppressed at order $\alpha_S$. 
For these processes conventional mesons
and four-quark states 
are only produced at order $\alpha^2_S$. Diffractive production prefers 
hybrid mesons, with glueballs, conventional mesons and four-quark states
suppressed at order $\alpha_S$.
The last four processes are glue-averse (glueballs at order $\alpha_S^2$,
and hybrids at order $\alpha_S$), and prefer conventional meson
production at order 1. Four-quark state production is order $\alpha^2_S$,
except for two-photon production at order 1.

The na\"{\i}ve power counting in Fig. \ref{prod} corresponds narrowly to
whether glueballs and hybrids are actually observed experimentally.

\begin{figure}
\begin{center}
\leavevmode
\hbox{\epsfxsize=5 in}
\epsfbox{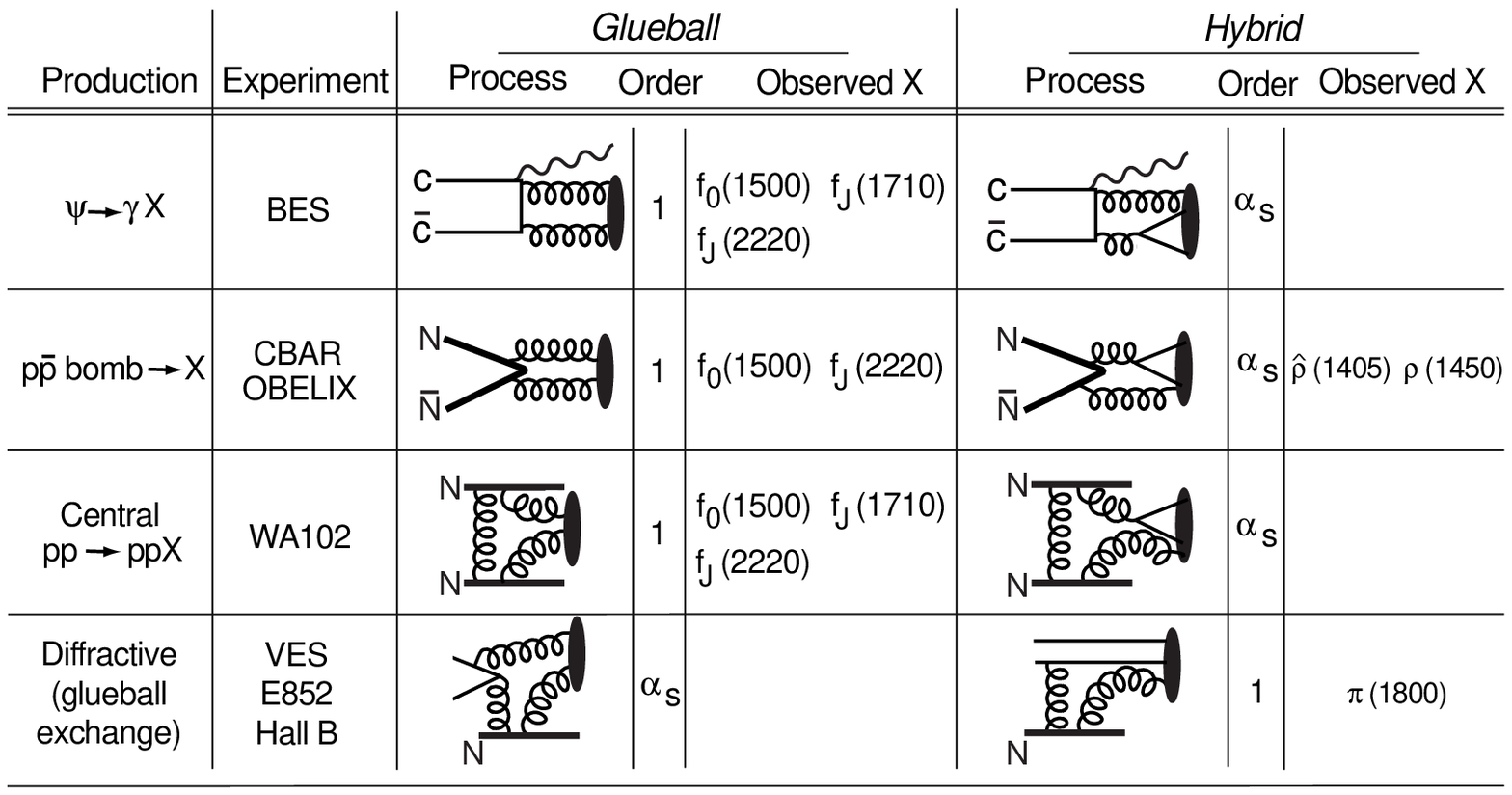}
\epsfbox{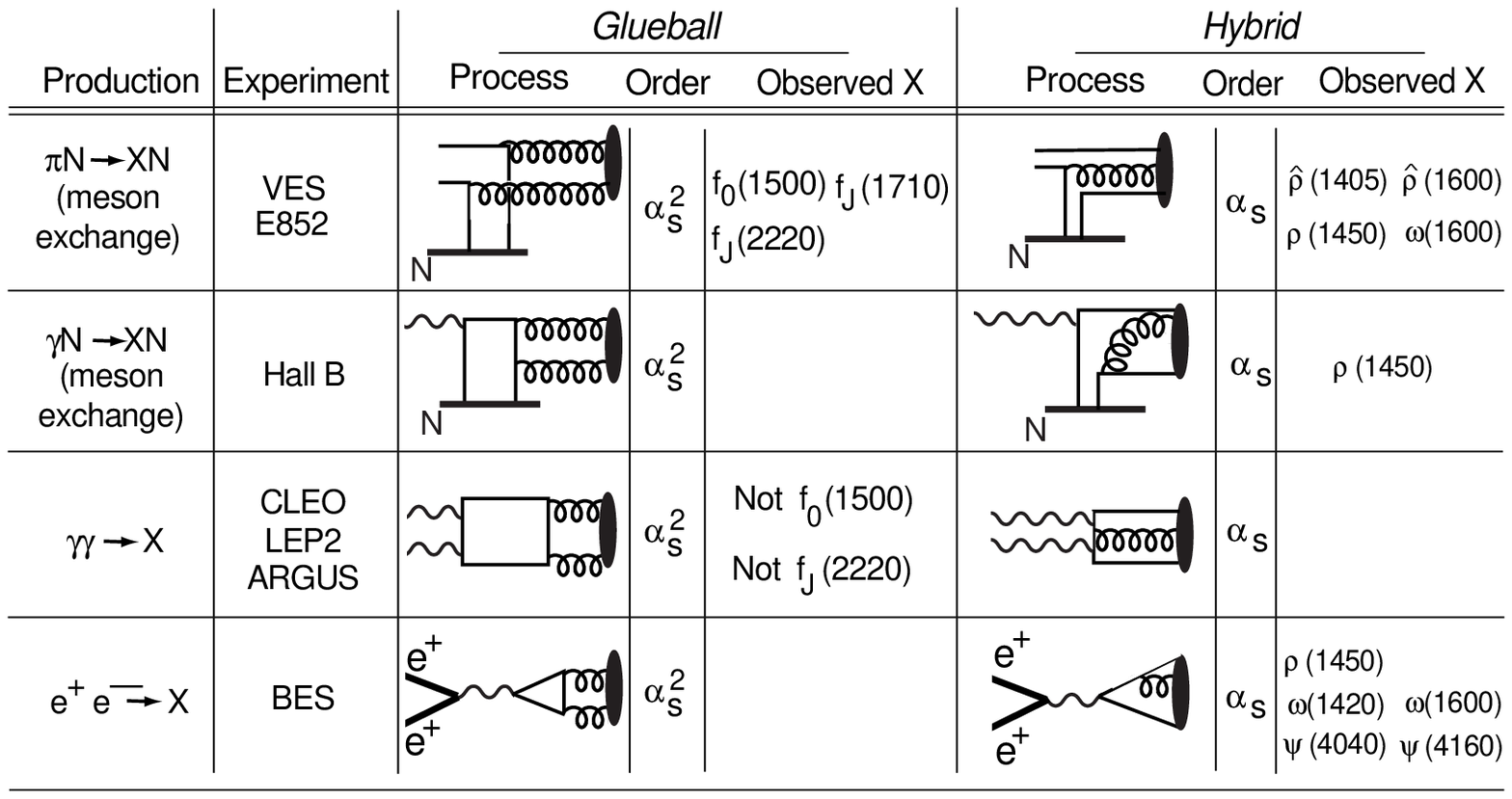}
\vspace{-1.2cm}
\end{center}
\caption{\plabel{prod}
Production processes. The nucleon $N$ is a proton or neutron.}
\end{figure}

\section{Pivotal Experimental Results}

The last decade marked the discovery of gluonic excitations, overturning
the traditional taxonomy of all known hadrons
as being either conventional mesons or baryons. The
$J^{PC} = 0^{++}$ (scalar) glueball has been discovered, although its
exact location in the spectrum has not yet been pinned down.
This can be regarded as the only robust experimental result on gluonic
excitations. Closely following is strong evidence for the existence of
two $1^{-+}$ exotic $I=1$ (isovector) states, something that could not
be said a decade earlier. Three 
issues of significant current interest will not be
covered: the ephemeral $2^{++}$ (tensor) cousin
of the scalar glueball, the possible four-quark nature of the $f_0(980)$ and
$a_0(980)$ and the existence or non-existence of a broad $\sigma$
resonance. Further
information on these subjects, and other outstanding puzzles of
hadron spectroscopy, can be found in detailed reviews \cite{godfrey99,amsler98}
. The search for hybrid baryons and four-quark states
is still a nascent field, reviewed in refs. \cite{barnes93,capstick00,godfrey99}.

\subsection{Scalar Glueball}

Significant advances have been made in clarifying the spectrum of
$J^{PC}=0^{++}$  $I=0$ (isoscalar) states. The $f_0(980)$  and
$f_0(1500)$ are today
the best established scalar isoscalar states. 
The subscript is the total angular momentum $J$, and the argument the
mass in MeV. 
Recently, clear
evidence for $f_0(1370)$ has emerged and a number of analyses are
converging on the $J=0$ assignment for $f_J(1710)$. There is a possible
higher mass resonance, or resonances, $f_0(2000-2100)$. Debate is still
raging about whether the low mass $\sigma$ phenomenon is
resonant or not. Details can be found in 
refs. \cite{amsler98,meyer99,amsler00}.

That the scalar resonances are too fecund is illustrated by the fact that
the Isgur-Godfrey relativized quark model expects only two scalar resonances
below 1.7 GeV, while probably more than two states are below this mass
(Figs. 16 and 22 of ref. \cite{godfrey99}).
This suggests the possibility of additional four-quark or
glueball states. A small subset of models does allow hybrid mesons in the
correct mass range, but we shall exclude this possibility.

The argument for the presence of a glueball amongst the scalar states
is firstly based on the convergence of lattice calculations on a primitive
glueball mass of around $1.6\pm 0.1$ GeV \cite{godfrey99}.
Note the proximity to $f_0(1500)$ and $f_J(1710)$.
Secondly, glueball character is indicated by production in
glue-rich processes and non-production in glue-averse processes, as
well as the so-called Close-Kirk filter, as we shall now elaborate.

The $f_0(980),\; f_0(1370),\; f_0(1500)$ and $f_0(1710)$  are strongly
produced in central production, where there are techniques to accertain
that they are produced mostly via the glue-rich collision
indicated in Fig. \ref{prod}. The two gluons connecting to the proton
are called the {\it pomeron}, so that the process can be thought of as
a double pomeron collision.
Also, $f_0(980), f_0(1370)$ and $f_0(1500)$ are strongly produced
in glue-rich $p\bar{p}$ annihilation. The $f_J(1710)$ is often at the edge
of phase space in $p\bar{p}$ annihilation,
so that its non-observation need not be significant.
Glue-rich $\psi$ radiative decay also significantly produces
$f_0(1500)$, $f_J(1710)$ and $f_0(2000-2100)$.
Close, Farrar and Li have developed quantitive techniques to extract
the gluon affinity for a state from $\psi$ radiative decay data. These
techniques indicate that $f_0(1500)$ and $f_0(1710)$ have substantial
glueball components.

Detailed analyses of the (mostly) double pomeron exchange process have been
performed. Consider the two-dimensional momentum vectors ${\bf p}_T^1$ and
${\bf p}_T^2$ for the two pomerons, where ``T'' indicates that the vectors
are the transverse components to the beam pipe. Define the magnitude
$dp_T \equiv |{\bf p}_T^1 - {\bf p}_T^2|$.
Grouping together resonances according to their $dp_T$ behaviour
yields that the
$f_0(980),\; f_0(1500)$ and $f_0(1710)$ behaves in the
opposite way to all well-established conventional mesons. The observation
 that
all conventional mesons behave in the same way is called the
{\it Close-Kirk filter}. Also, $f_0(1370)$ 
has a behaviour somewhere between conventional mesons and
$f_0(980),\; f_0(1500)$ and $f_0(1710)$.
The errant behaviour of the isoscalar scalar states is
taken to mean that they  contain something beyond conventional mesons.
The higher mass $f_0(2000-2100)$ behaves like a conventional meson.

It is also instructive to look at the non-appearance of states in
glue-averse two photon production. The ALEPH collaboration at LEP2 provided
a restrictive bound on the two-photon width of the $f_0(1500)$.
The $f_0(980)$ also has a small two-photon width
\cite{pdg98}.
On the other hand, $f_0(1370)$ has a
two-photon width $5.4\pm 2.3$ keV \cite{pdg98} which is perfectly consistent with
expectations for a conventional $n\bar{n}\equiv \frac{1}{\sqrt{2}}(u\bar{u}+
d\bar{d})$ meson. There is currently no definitive measurement for
$f_J(1710)$ where $J$ has been determined.
However, as we shall see below, it is possible to explain the small two-photon
width of $f_0(1500)$ without invoking a glueball.

It is clear that production processes indicate that the primitive
glueball might be distributed over more than one physical state:
notibly $f_0(1500)$ and $f_J(1710)$.
This implies that there is significant mixing between primitive glueballs and
mesons. 

We now analyse the mixing mathematically. Assume
that a glueball couples to a pair of primitive mesons, one with
flavour $n\bar{n}$ and the other with flavour $s\bar{s}$. The coupling
 to intermediate decay channels is neglected for the purposes of this
introductory orientation. 
Then, we have the
following $3\times 3$ hermitian mass matrix, where, in addition to the meson
mixing amplitude $A$, we have the amplitude for glueball-meson mixing
which we denote by $g$:
\begin{equation}
\plabel{mat}
{\cal M}=\left(
\begin{array}{ccc}
G & g & gr \\
g^\ast  & S+A & Ar \\
g^\ast r^\ast  & Ar^\ast  & N+A|r|^2
\end{array}
\right) ,
\end{equation}
where $G,S,N$ indicates the (real) 
primitive masses. Here
$\langle G | n\bar{n} \rangle = r\; \langle G | s\bar{s} \rangle$ and
$\langle n\bar{n} | n\bar{n} \rangle = r\;  \langle n\bar{n} | 
s\bar{s} \rangle = |r|^2\;  \langle s\bar{s} | s\bar{s} \rangle$.
In the $SU(3)$ limit one can
use the methods of Eq. \ref{gdec} to show that $r=\sqrt{2}$.

Note that $A$ must be real for the matrix to be hermitean. With $g$ and
$r$ both real the matrix is the most general parametrization
of $3\times 3$ (real) symmetric matrix, since it contains six independent
parameters. 

The matrix is diagonalized
${\cal M}\Longrightarrow {\rm diag}\left( \tilde{G}, \tilde{ S},\tilde{N}\right)$ by the masses of the three physical
states, which are 
determined from the
three eigenvalue ($\lambda$) 
equations (which follow from ${\rm Det}\left( {\cal 
M}-\lambda I\right) =0).$ Eliminating $A$ and $g$
from the eigenvalue equations leads, upon some algebra, to the formula

\[\Big[ (1+|r|^2)\;\!\tilde{G}-|r|^2\;\!S -N\Big]
\Big[ (1+|r|^2)\;\!\tilde{S}-|r|^2\;\!S-N
\Big] \Big[ (1+|r|^2)\;\!\tilde{N}-|r|^2\;\!S
-N\Big]\]

\vspace{-.8cm}\begin{equation}
+\;\!\Big[ (1+|r|^2)\;\!G-|r|^2\;\!S-N\Big] \;|r|^2\;\!
\Big[ S-N\Big] ^2=0 \plabel{gsch}
\end{equation}
which is called the {\it generalized Schwinger mass formula}.
In 1964 Julian Schwinger derived a 
simpler, phenomenologically successful, formula for the case where
there is no glueball.
His matrix is just the right-bottom $2\times 2$ sub-matrix of the
$3\times 3$ matrix (Eq. \ref{mat}), restricted to be real
 with $r=\sqrt{2}$. We note that the generalized Schwinger mass
formula does not depend on the couplings $A$ or $g$. This is
very useful, as they are difficult to extract from experiment.

Assume that there is no direct coupling between mesons,  i.e. that $A=0$. 
The 
coupling between mesons $A$ can be shown to be
suppressed as $\frac{1}{\sqrt{N_c}}$ relative to the
glueball-meson coupling $g$, where $N_c$ is the number of
colours in QCD \cite{lebed99}. This result comes from the first level of
understanding. When  $A=0$
one can combine Eq. \ref{gsch} 
with the trace condition for the matrix (Eq. \ref{mat}),
\begin{equation}
\tilde{G}+\tilde{S}+\tilde{N}=G+S+N
\end{equation}
in order to determine $two$ unknown masses. The strategy is to
assume a value for $r$, and use four input masses to predict
the remaining two masses.

Now specialize to real and positive $g$.
Once all primitive and physical masses are known there are formulae
that enable calculation of the coupling constant $g$, as well as the
matrix that diagonalizes $\cal M$, called the {\it valence content}
matrix. This matrix contains the eigenvectors of $\cal M$ in its
columns. The formulae are now exhibited without proof.

The coupling can be calculated from the masses according to

\beqn\plabel{gg}
\sqrt{-\frac{(S-\tilde{G})(S-\tilde{S})(S-\tilde{N})}{S-N}} = g =
\sqrt{-\frac{(N-\tilde{G})(N-\tilde{S})(N-\tilde{N})}{r^2(N-S)}}
\eeqn

If we write the valence content of the physical state
$X$, either the physical glueball, $s\bar{s}$ or $n\bar{n}$,
as $|X\rangle  = X_G |G\rangle + X_S |S\rangle + X_N |N\rangle$,
one requires that $|X\rangle$ be normalized, i.e. that
$X_G^2+X_S^2+X_N^2 = 1$. It is possible to show that the
valence content can be explicitly calculated as:

\beqn\plabel{val}
X_G = {\cal N}_X \hspace{1cm} X_S = {\cal N}_X \frac{g}{\tilde{X}-S}\hspace{1cm}
X_N = {\cal N}_X  \frac{r g}{\tilde{X}-N}
\eeqn
where $\tilde{X}$ is the physical mass of state $X$ and
\beqn\plabel{nn}
\frac{1}{{\cal N}_X} = \sqrt{1+(\frac{g}{\tilde{X}-S})^2 + (\frac{r g}{\tilde{X}-N})^2}
\eeqn

Note that the valence contents are only specified up to an overall
sign, i.e. one cannot distinguish between 
$X_G,X_S,X_N$ and $-X_G,-X_S,-X_N$. Eqs. \ref{gg} - \ref{nn} have been 
checked numerically.

We shall now consider four limiting scenarios, and study the valence
content of the physical glueball in each case, taking $r = \sqrt{2}$
for simplicity.

{\bf $\boldmath SU(3)$ symmetry}: This arises in two cases. 

First take the 
$SU(3)$ limit $S=N$ and $r=\sqrt{2}$. From Eq. \ref{val} this implies that 
$X_S : X_N = 1: \sqrt{2}$, i.e. that the physical glueball has
flavour content proportional to $u\bar{u}+d\bar{d}+s\bar{s}$. This is 
an $SU(3)$ {\it singlet}: Since the primitive glueball carries no 
flavour, i.e. is an $SU(3)$ singlet, we expect that it should only mix with the
$SU(3)$ singlet quark flavour combination.

Secondly consider a physical glueball much higher in mass than the
primitive $s\bar{s}$ and $n\bar{n}$.
Again $X_S : X_N = 1: \sqrt{2}$, i.e. the physical 	glueball has 
the same flavour content as before.

{\bf Midway}: Consider a physical glueball halfway between the 
primitive $s\bar{s}$ and $n\bar{n}$ states. 
Then $X_S : X_N = 1: -\sqrt{2}$, i.e. 
the physical glueball has
flavour content proportional to $u\bar{u}+d\bar{d}-s\bar{s}$. This is 
somewhere between the {\it ideal} mixing assignment $u\bar{u}+d\bar{d}$ and the
$SU(3)$ {\it octet}
 $u\bar{u}+d\bar{d}- 2 s\bar{s}$, and is, perfunctorily,  a popular choice for
the flavour content of the $\eta$.
 
{\bf $\boldmath SU(3)$ Octet}: With the physical glueball between the 
primitive $s\bar{s}$ and $n\bar{n}$ states, but two times further from 
the $n\bar{n}$ than from the $s\bar{s}$, one obtains 
flavour structure of the physical glueball proportional to
$u\bar{u}+d\bar{d} - 2 s\bar{s}$. This is an $SU(3)$ octet, indicating that
$SU(3)$ symmetry is maximally violated.   

{\bf $\boldmath n\bar{n}$}: When the physical glueball mass is near the 
primitive $n\bar{n}$ mass, $X_G : X_S : X_N = 0 : 0 : 1$. This means
that the physical glueball undergoes very strong mixing and becomes the
primitive $n\bar{n}$!

{\bf $\boldmath s\bar{s}$}: Similarly, 
for a physical glueball near the primitive
$s\bar{s}$ mass,  $X_G : X_S : X_N = 0 : 1 : 0$, so that the physical
glueball is just the primitive $s\bar{s}$. 

It is clear that one can consider the physical glueball at various places
between the primitive $n\bar{n}$ and $s\bar{s}$, and obtain any desired
ratio $X_S : X_N$ with the restriction that the sign of  $X_S$ and
$X_N$ is different.

On the other hand, if the physical glueball is either above or
below {\it both}
the $n\bar{n}$ and $s\bar{s}$ states,  the sign of $X_S$ and $X_N$ 
will be the same.

\begin{table}[t]\centering
\begin{tabular}{|c|c|c|c|c|c|c|c|c|c|c|} \hline
& \multicolumn{2}{c|} {$SU(3)$ singlet} & \multicolumn 
{2}{c|}{Midway} & \multicolumn
{2}{c|}{$SU(3)$ Octet} & \multicolumn {2}{c|}{$n\bar{n}$} & \multicolumn
{2}{c|}{$s\bar{s}$}
\\ \cline {2-11}
&  \ \ \ $\cal A$ \ \ \ & $\Gamma$ & $\cal A$ & $\Gamma$ &  \ \ \ $\cal A$ \ \ \ & $\Gamma$ & $\cal A$ & $\Gamma$ & $\cal A$ & $\Gamma$  \\ 
\cline {2-11}
$\pi \pi $ & 1 & 3 & 3 & 27& 2 & 3 & 6 & 27 & 0 & 0   \\
$K \bar{K} $ & 1 & 4 & 0 & 0 & $-1$ & 1 & 3 & 9 & 3  & 9   \\
$\eta \eta $ & 1 & 1 & $-1$& 1 & $-2$ & 1 & 2 & 1 & 4  & 4   \\
$\eta^\prime \eta $ & 0 & 0 &$2\sqrt{2}$& 16 &$2\sqrt{2}$  & 4 & $2\sqrt{2}$ & 4 & $-2\sqrt{2}$ & 4   \\
\hline
\end{tabular}
\caption{\plabel{dec}
Amplitude $\cal A$ and width $\Gamma$ ratios of a physical glueball decaying to pseudoscalar final states.}
\end{table}

Now consider decays.
Assume that the primitive $n\bar{n}$ and $s\bar{s}$ decay
to $\pi\pi,\; K\bar{K},\; \eta\eta$ and $\eta'\eta$ via connected
decay. Also assume that the primitive glueball does not
decay to these final states, i.e. its decays are subdominant as
expected from the OZI rule. If the total decay width is below expectations
for conventional mesons,
as is the case for the $f_0(1500)$ and $f_J(1710)$, that may indicate
a substantial glueball valence content. However, the total decay widths
are not small, as expected for an unmixed glueball.

The decay of the physical glueball can be calculated by considering
the decay of its primitive $n\bar{n}$ and $s\bar{s}$ valence content
(see problem \ref{sch}). The
amplitudes are obtained in the same manner as Eq. \ref{gdec}, yielding

\begin{eqnarray}
\left <n\bar{n}|\pi^+\pi^- \right > = \left < n\bar{n}|\pi^0 \pi^0 \right >
= \sqrt{2} & &   
\left <s\bar{s}|\pi^+\pi^- \right > = \left < s\bar{s}|\pi^0 \pi^0 \right >
= 0        
\nonumber\\
\left <n\bar{n}| K^+ K^- \right > = \left <n\bar{n}| K^0 \bar{K}^0 \right > 
= \frac{1}{\sqrt{2}}  & &  
\left <s\bar{s}| K^+ K^- \right > = \left <s\bar{s}| K^0 \bar{K}^0 \right > 
= 1                  
\nonumber\\
\left <n\bar{n}|\eta \eta \right >  = \frac{\sqrt{2}}{3} & & 
\left <s\bar{s}|\eta \eta \right >  = \frac{4}{3} 
\nonumber\\
\left < n\bar{n}| \eta' \eta \right >  = \frac{2}{3} & & 
\left < s\bar{s}| \eta' \eta \right >  =-\frac{2\sqrt{2}}{3}  
\end{eqnarray}

\begin{table}[t]\centering
\begin{tabular}{|c|c|c|c|c|c|} \hline
& { $\sigma\sigma$} & { $\rho\rho$} & { $\pi(1300)\pi$} &
   { $a_{1}(1260)\pi$} & {  Sum$_1$} \\ \hline
{ $f_{0}(1370)$}  & { $105.2\pm 32.0$} & { $76.8\pm 37.0$} &
{ $6.6\pm 4.2$} & { $37.2.0\pm 16.3$} & { $226$} \\
{ $f_{0}(1500)$}  & { $15.6\pm 9.2$} & { $6.5\pm 5.9$} &
{ $9.8\pm 7.7$} & { $7.9\pm 5.5$} & { $40$} \\\hline
  & { $\pi\pi$} & { $\eta\eta$} & { $\eta\eta'$} & 
{ $K\bar{K}$} &
    { Sum$_2$} \\\hline
{ $f_{0}(1370)$}  & { $19.2\pm 7.2$} & { $0.4\pm 0.2$} &
                         & { $7.0\pm 1.6$ } & { $32.5$} \\
  & & & & { $18.8\pm 4.0$} & \\
{ $f_{0}(1500)$}  & { $24.6\pm 2.7$} & { $1.91\pm0.24$} &
{ $1.61\pm 0.06$} & { $4.52\pm 0.36$} & { $32.6$} \\
\hline
\end{tabular}
\caption{\plabel{cbar}
Crystal Barrel widths in MeV, ca. 2000.  }
\end{table}

The amplitudes and
widths are displayed in Table \ref{dec}, up to an arbitrary normalization. It
is evident that predictions for the widths vary widely, indicating the
sensitivity of experimental widths to the valence content of the state.
The reader is invited to determine which pattern of widths best correspond
to the most recent experimental data from Crystal Barrel in Table \ref{cbar}.
It is clear that the data for $f_0(1500)$ are not consistent
with the $SU(3)$ singlet / unmixed glueball (flavour democratic) or the
$s\bar{s}$ interpretation. The small two-photon width of $f_0(1500)$ 
also excludes the $n\bar{n}$ interpretation.
However, there exists a valence content for the physical glueball that
gives zero two-photon width (see problem \ref{pho}).
The above argues that mixing is needed to explain the decay pattern of
$f_0(1500)$.

	An ingredient which is currently missing in the description of the
scalar isoscalar states is the consideration of, amongst others,
 radially excited quark model states. This would lead to at least
$5\times5$ matrices. These, and higher dimensional 
analogues are known to obey generalized Schwinger formulae
and a trace condition similar to the ones derived in this
section.

 Does the substantial scalar glueball-meson mixing 
imply that the same is true for glueballs with other $J^{PC}$?
I.e. are other glueballs also not narrow and hence difficult to detect
experimentally?
It is clear that the higher the primitive glueball mass, the more
conventional and hybrid mesons will have similar masses, since there is
both a tower of radially and orbitally excited states,
and a tower of different types of hybrid mesons. Can 
glueball-meson mixing be suppressed?
There is currently no theoretical consensus on this issue.


\subsection{Isovector $J^{PC}=1^{-+}$ Exotics}

Evidence for an embarrassment of riches of
 isovector $J^{PC}=1^{-+}$ exotic enhancements 
$\hat{\rho}(1405)$ at mass $1392^{+25}_{-22}$ MeV, width
$333\pm 50$ MeV \cite{pdg98} and $\hat{\rho}(1600)$ at 
mass $1593\pm 8$ MeV, width $168\pm 20$ MeV has recently emerged
\cite{godfrey99}.
The former enhancement  is observed by both E852 and Crystal Barrel 
in very different production processes decaying
the $\eta\pi$. The 
enhancement $\hat{\rho}(1600)$ was observed by E852  decaying to
$\rho\pi$. There is also some weaker evidence from E852 and VES
that it decays to $\eta'\pi$ and $b_1\pi$, but {\it not} to $\eta\pi$
and $f_2\pi$.
Evidence for higher mass states is more tentative.

In experimental analyses the observed enhancements are described by  complex
amplitudes, with both a magnitude and a phase. The change of the phase
as one moves from low to high four-momentum squared 
$(p_B+p_C)^2$ of the final decay
channel, e.g. $\eta\pi$, is called {\it phase motion}. Here
$p_B$ and $p_C$ denote the four-momenta of the final states.
The phase motion is expected to go through $180^o$
for a simple resonance. This enables one to determine 
whether the observed enhancements are
resonant or not. Let's take the $\hat{\rho}(1405)$ as an example. 
Crystal Barrel recently claimed that the phase motion
in the $\eta\pi$ P-wave goes through $213^o \pm 5^o$, consistent with
expectations for a resonance. 
At E852 there is a well-known resonance
$a_2$ decaying to $\eta\pi$ which dominates the $\hat{\rho}(1405)$.
This raises the prospect that experimental misidentification might lead
to the $a_2$ appearing in the $J^{PC}=1^{-+}$ amplitude. It is 
frequently argued that this circumstance would lead to a fake
$J^{PC}=1^{-+}$ amplitude having the same phase motion as the
$a_2$ amplitude. This is based on the idea that experimental 
misidentification {\it cannot} by itself lead to phase motion. 
If one studies the relative phase motion between the $J^{PC}=1^{-+}$ 
and $a_2$ amplitudes, and finds this to be constant, one should therefore
conclude that the $J^{PC}=1^{-+}$ amplitude is due to experimental 
misidentification. E852 did not observe this constancy, and hence concluded 
that the $\hat{\rho}(1405)$ was resonant. This interpretation might
be overly simplistic in view of the fact that there is still the possibility
of non-resonant $\eta\pi$ production, which can interfere with a 
resonant $\hat{\rho}(1600)$, and appear as an apparent 
resonance at the mass of the $\hat{\rho}(1405)$. 
This mechanism can in fact account for the
E852 data \cite{godfrey99}. However, from Occam's razor
and the independent Crystal Barrel observation, I shall be 
predisposed towards the simpler E852 interpretation for the
remainder of this lecture.

Phase motion of the $\hat{\rho}(1600)$ against $\pi(1300)/\pi(1800)$,
$a_1$, $a_2$ and $\pi_2(1670)$ has also been
observed by E852, and was interpreted as
evidence for the resonant nature of the enhancement.

Manifestly exotic $J^{PC}$ isovector quantum numbers immediately
translate into either a hybrid meson or four-quark interpretation 
for the resonances. 

Since $\hat{\rho}(1405)$ has only been observed in $\eta\pi$, it is natural
to assume that the decay has a substantial branching ratio. If this
is the case, the observed decay is in contravention with 
selection rule I of section \ref{selsec}. This means that either the
OZI rule is violated or that $\hat{\rho}(1405)$ is not dominantly a
hybrid meson. The latter would in itself be an important result, signalling 
the observation of a four-quark state. The only other decay
channels with substantial phase space are $\rho\pi$ and $\eta^{'}\pi$.

The $\hat{\rho}(1600)$ has enough phase space to decay to
$K^{\ast}K,\;  b_1\pi,\; 
f_1\pi,\; f_2\pi$ and $\eta(1295)\pi$ in addition. Selection
rule II of section \ref{selsec} appears to say that decay
should only be to non - $L_B=L_C =0$ mesons, i.e. to 
$b_1\pi,\;  f_1\pi,\;  f_2\pi$ and $\eta(1295)\pi$. This observation has
important experimental consequences, as $b_1,\; f_1,\; f_2$ and $\eta(1295)$
decay on a strong interaction time scale to other particles, so that
the final state is complicated. This stands in marked contrast the final
states $\eta\pi,\; \eta^{'}\pi$ and $K^{\ast}K$, where 
$\eta,\; \eta^{'}$ and $K^{\ast}$ are almost
stable on the strong interaction time scale.

Selection rule II only holds if the final states can be regarded as the same,
except for their flavour and spin. For example, in the decay to
$\rho\pi$, the $\rho$ and $\pi$ clearly have different flavours and
spins. This does not break the selection rule. However, $\rho$ and $\pi$
have different sizes, which does break the selection rule. Hence the
selection rule is not exact.
VES quotes the width ratios $\rho\pi: \eta'\pi : b_1\pi
= 1.6 \pm 0.4 : 1.0 \pm 0.3 : 1 $ for $\hat{\rho}(1600)$.
E852 sees $\hat{\rho}(1600)$ in $\rho\pi$ and $\eta^{'}\pi$, but not in
$f_2\pi$.
This appears to challenge the validity of the selection rule
and hence current models which imply it \cite{swanson97}.

\section{No Conclusions}

The ideas presented here constitute some of the phenomenologist's 
language to describe experiment, incorporating ideas 
about gluon excitations from QCD.
It is possible that this whole beautiful structure
will be swept away by a thunderbolt from lattice QCD or experiment.

\vspace{.2cm}

This research is supported by the Department of Energy under contract
W-7405-ENG-36. Useful discussions with F.E. Close and C.A. Meyer are
 gratefully
acknowledged.

\section{Problems}

\newtheorem{problem}{}

{\bf Epicurean}:

\begin{problem}
\plabel{hbar} {\em The hybrid meson candidate $\pi(1800)$ is strongly
produced
in diffractive $\pi N$ collisions. Based on this,
which production
process discussed is expected to copiously produce the $N\frac{1}{2}^+$ 
hybrid baryon?}
\end{problem}

\begin{problem}
\plabel{psi4040} {\em Assume that the $e^-e^+$ widths of $\psi(4040)$ and 
$\psi(4160)$ are approximately the same, and that hybrid mesons have
negligible $e^- e^+$ widths. Explain how the two physical states can
be constituted from a primitive conventional and hybrid meson. Why should
the decay pattern to other final states of $\psi(4040)$ and 
$\psi(4160)$ be closely related?}
\end{problem}

\begin{problem}
\plabel{singlet} {\em Can only glueballs decay flavour democratically?
(Refer to Tables \ref{demo} and \ref{dec}).}
\end{problem}

\begin{problem}
\plabel{pho} {\em Take into account that the $u,d$ and $s$ quarks respectively
 have electric charges $\frac{2}{3},-\frac{1}{3}$ and $-\frac{1}{3}$, and
assume lowest order electromagnetic coupling of quarks and vanishing
 electromagnetic coupling of gluons. Show that for $X_S : X_N =
5 : -\sqrt{2}$ the two-photon decay of the physical glueball vanishes in 
the $SU(3)$ limit.}
\end{problem}

\noindent {\bf Stoic}:

\begin{problem}
\plabel{qex} {\em By considering that a conventional 
baryon has two independent quark positions
 in its centre of mass frame, argue that the combination
$L^P = 0^-$ is the only $L^P$ combination that 
cannot be constructed for baryons in the non-relativistic
quark model, i.e. that it is ``quark model exotic''. Here $L$ is the total
orbital angular momemtum of the baryon.}
\end{problem}

\begin{problem}
\plabel{flux} {\em Consider the connected 
decay of an adiabatic hybrid meson $A$ with
$\Lambda = \Lambda_A$ to  two adiabatic hybrid mesons $B$ and $C$ with 
$\Lambda = \Lambda_B$ and $\Lambda_C$ respectively. 
Denote the quark-antiquark
line of $A$ by $\hat{\br}$. Decompose
the pair creation position $\bf y$ from the
midpoint of the quark-antiquark line of $A$ in polar coordinates.
Defining $\phi$ to be the angle of $\bf y$ around the $\hat{\br}$-axis derive
the following result related to the conservation of
angular momentum around the $\hat{\br}$-axis:
The most general form of the flux-tube overlap in the limit where 
pair creation is near to the initial quark-antiquark line is
proportional to $e^{i (\Lambda_{A} - \Lambda_{B} - \Lambda_{C}) \phi}$. }
\end{problem}

\begin{problem}
\plabel{ssbar} {\em Why are $s\bar{s}$ excited conventional 
mesons rarely seen in production processes studied experimentally?
Specifically, why are they suppressed in 
central production?}
\end{problem}

\noindent {\bf Herculean}:

\begin{problem}
\plabel{ccbar} {\em List the decays of a $J^{PC} = 0^{+-}$ $c\bar{c}$ exotic
below the $D^{\ast\ast}D$ threshold, where $D^{\ast\ast}$ denotes the
$L=1$ conventional charm-light mesons. Argue that decays to
$D\bar{D},\; D^{\ast}\bar{D}$ and $D^{\ast}D^{\ast}$ are forbidden.
Which decay mode should $0^{+-}$ be searched in?}
\end{problem}

\begin{problem}
\plabel{symm} {\em Show that the gluons in adiabatic hybrid mesons with a 
fixed quark and antiquark are characterized by 
three conserved quantum numbers:
(1) The magnitude $|\Lambda|$ 
of the angular momentum of the gluons projected onto the
quark-antiquark line, (2) $CP$ around the midpoint between the
quark and the antiquark, and (3) if $|\Lambda|=0$, 
reflection in the plane containing the
quark-antiquark line.}
\end{problem}

\begin{problem}
\plabel{sch} {\em Rewrite Eq. \ref{mat} as a hamiltonian quadratic in the
fields corresponding to the primitive states. Introduce an additional term
which describes the coupling of each primitive state to a specific  decay
channel. Now make a transformation from primitive fields to physical 
fields, equivalent to diagonalizing Eq. \ref{mat}. Note that the
unitary matrix that attains this is the valence 
content matrix. Show that in order to calculate the decay amplitude of
a physical state to the decay channel, it is necessary to add the 
decay amplitudes of all its primitive states, weighted by their
valence content.}
\end{problem}

\end{document}